# Correlation of the dynamic contact angle with the capillary number and its hysteresis

Jun Li

**Abstract**

The moving contact-line problem is of both theoretical and practical interest. The dynamic contact angle changes with the capillary number defined by the contact-line speed, and the correlation also depends on the equilibrium contact angle measured at the static state. This correlation is usually required as an input to the traditional solvers based on the Navier-Stokes-like equations, but it is simulated as an output in the current study using the lattice Boltzmann method (LBM) in a displacement process of two-immiscible fluids. The macroscopic theory and the molecular dynamics (MD) simulation had shown a linear scaling law for the cosine of dynamic contact angle, which is also observed in the previous LBM study in a short range of small capillary numbers and for two neutral wetting conditions. However, our study shows that this linear scaling law holds in the whole range of capillary numbers and is universal for all wetting conditions. In a special case of complete wetting (spreading) with a zero equilibrium contact angle, a thin film of the wetting fluid occurs when the wettability is very strong, which leads to a hysteresis that substantial capillary number is required to initiate the deviation of the dynamic contact angle from its equilibrium state. This observation is consistent with the previous report on a new mechanism for the static contact angle hysteresis due to the presence of free liquid films. With an increasing capillary number, the fluid-fluid interface starts oscillating before fingering. Different fingering patterns are observed for cases with different equilibrium contact angles.

1. Introduction

In the traditional study on contact angle, a sessile droplet is usually placed on a substrate. When additional fluid is added to the droplet, the contact line advances and the droplet exhibits an advancing contact angle. Alternatively, if fluid is removed from the droplet, the contact angle first decreases to a receding value before the contact line retreats. For these two processes, the contact angles measured after we stop adding or removing fluid (i.e., the contact-line speed becomes zero) are different and the difference is referred to as the contact angle hysteresis. Hysteresis can arise from surface anomalies, like roughness or heterogeneity [1-3], because the contact line could be pinned when it reaches the edge of an asperity. This hysteresis can also occur on smooth, homogeneous surfaces due to the presence of free liquid films, which can be interpretated using surface forces [4].

If the interface keeps on moving, the advancing or receding contact angle changes with the contact-line speed. Many simulation studies have been conducted using the lattice Boltzmann method (LBM) [5-11]. This correlation is required as an input in constructing empirical boundary conditions for modelling problems of capillary hydrodynamics [12] if the traditional computational fluid dynamics (CFD)

Email: lijun04@gmail.com

solvers based on the Navier-Stokes-like equations are employed. In contrast, this property can be automatically captured by the LBM simulation as an output.

Like the droplet movement, the displacement of a fluid by another immiscible fluid in a capillary tube is also ubiquitous. Several models are proposed to describe the dynamic (only advancing here) contact angle measured through the injected fluid as a function of the capillary number $Ca$ and the static advancing contact angle [13-15]. The model coefficients need to change with the range of capillary number in order to match with the experimental data [15]. This correlation is of both fundamental [16-17] and practical [18] interest. In general cases, the dynamic contact angle is not only a function of the contact line speed but also a function of the flow field in the vicinity of the moving contact line [12, 17, 19, 20]. At large $Ca$, the interface becomes unstable and starts fingering. Due to the symmetric property, the fingering pattern will be symmetric, which is different from the fingering of a moving droplet [7-11, 21]. We simulate this displacement process using the LBM. As the surface in our simulations is ideally smooth and chemically homogenous, the contact angle at static state will be referred to as the equilibrium contact angle, as in [4], that is reached by an automatic force balancing process starting with a flat interface. Our study focuses on the correlation of dynamic contact angle with $Ca$ and the equilibrium contact angle. Meanwhile, the evolution of a stable interface shape with $Ca$ and the final growing finger at large $Ca$ are discussed to provide more insights into this correlation.

## 2. Problem statement

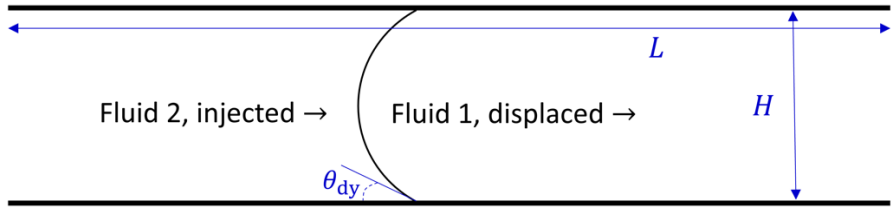

Fig. 1 Schematic of the two-phase channel flow for the study of dynamic contact angle.

As shown in Fig. 1, the problem of injecting one fluid to displace another immiscible fluid through a 2D channel will be investigated. When the interface starts moving, the dynamic contact angle $\theta_{\text{dy}}$ will deviate from its equilibrium value $\theta_{\text{eq}}$ measured at static state. At different moving speeds (or $Ca$ in general), we will have either a stabilized interface, from which $\theta_{\text{dy}}$ can be measured, or a growing finger. Accordingly, there is a critical $Ca^*$, beyond which the interface starts fingering. The precise measurement of $Ca^*$ is challenging as we can only approach it with a continuously decreasing step. Instead, we will show the results for $Ca < Ca^*$ with a stable interface and $Ca > Ca^*$ with a growing finger such that $Ca^*$ can be estimated with the lower and upper bounds. For the cases with a stable interface, the correlation of $\theta_{\text{dy}}$ with $Ca$ and $\theta_{\text{eq}}$ will be studied. To make the definition and discussion clear, we limit the current investigation to flows in a 2D straight channel.

The capillary number is adjusted by the injecting speed while it is always defined by the actual moving speed $U$ of the contact line (or contact points in 2D cases), which is almost the same as the injecting speed in the cases with a stable interface. However, when fingering occurs, the fingertip moves faster than the contact line and the injecting speed lies in between since the finger continuously grows until having droplet pinch-off, which repeats with time.

### 3. LBM setting

Different from other options [22-24], the Shan-Chen multicomponent LBM model [25, 26] is used here to simulate the flow of two immiscible fluids because it complies with the physics of the Young equation by regulating the three interfacial tension coefficients to match the equilibrium contact angle $\theta_{eq}$ (i.e., wettability). Note that the validity of the Shan-Chen LBM in modeling the dynamic contact angle at different capillary numbers $Ca$ has been verified in [5]. Here we use the same algorithm to study this problem again but in the whole range of $Ca$ up to the critical value, after which the fingering occurs. Additionally, we consider different $\theta_{eq}$ ranging from 0 to $\approx \pi$ while $\theta_{eq}$ used in [5] is limited to $\pi/2$ and $0.4833\pi$, respectively. By doing so, we will investigate the behaviors of dynamic contact angle in a complete range of $Ca$ and $\theta_{eq}$.

To make it clear for applications, the quantities used here are dimensional with the SI base units and the implemented algorithm is summarized in Appendix. The lattice spatial distance is $\Delta x = 10^{-6}$ and the time step is $\Delta t = 10^{-7}$ making the lattice speed $c = 10$. At the initial state, the 2D channel is dominated by the fluid 2 for $x \leq 0.1L$ and by the fluid 1 for $x > 0.1L$, with a flat interface separating the two fluids. The dominant density $\rho_l$ and dissolved density $\rho_g$ for the initialization are 500 and 15, respectively, which are close to their equilibrium values when the interaction coefficient $G$ is set to 0.003 and the dimensionless relaxation times are $\tau_1 = \tau_2 = 0.9$. Here, the two fluids are not absolutely immiscible but the dissolved density is very small compared to the dominant density, which serves as a good approximation to the immiscible fluids. Note that the surface tension coefficient $\sigma$ and $\rho_l/\rho_g$ depends on $G(\rho_l + \rho_g)$ as discussed in [27] but this dependence changes with $\tau_1, \tau_2$ as observed in our simulations. Additionally, the equilibrium contact angle $\theta_{eq}$ depends mainly on the difference between effective wall densities $\rho_w^1, \rho_w^2$ of the two fluids [27] but also changes slightly with $\tau_1, \tau_2$. In our study, $\tau_1 = \tau_1 = 0.9$ is fixed and thus the two fluids have the same dynamic viscosity $\mu_1 = \mu_2 = 2/3000$. Additionally, $\sigma$ measured through Young-Laplace law is about $3.1 \times 10^{-3}$. The channel height $H$ is fixed at 0.0002 and discretized by 200 uniform grids, which are much more than 20 grids used in [5]. The channel length $L$ starts with 0.004, which corresponds to 4000 grids, and is increased to 0.008 or 0.015 for large injecting speeds such that the stabilized interface is still far away from the outlet to reduce the end effect.

### 4. Results and discussion
#### 4.1. Initialization and convergence processes

As shown in Fig. 2, the saturation $S$ of injected fluid becomes stabilized after the initial $500,000\Delta t$, after which the fluid 2 is injected from the left end. For the cases with a stable interface, the saturation increase rate is related to the speed $U$ of the contact-line movement after the interface shape is stabilized, which is indicated by a constant increase rate of $S(t)$. In the case of Fig. 2, we have $U = \frac{(S_2-S_1)L}{t_2-t_1} = \frac{(0.264363-0.191256)\times 0.008}{80000\times 10^{-9}}$ =7.311. If we directly monitor the interface locations at different moments, the computed speed of contact-line is about 7.325, which is consistent with the prediction of using $S(t)$. In the cases with a growing finger, $U$ will be directly computed from the variation of interface location with time. The insert of Fig. 2 also shows that the final interface shape is stabilized to allow the measurement of dynamic contact angle, which is noticeably different from its equilibrium value at the static state.

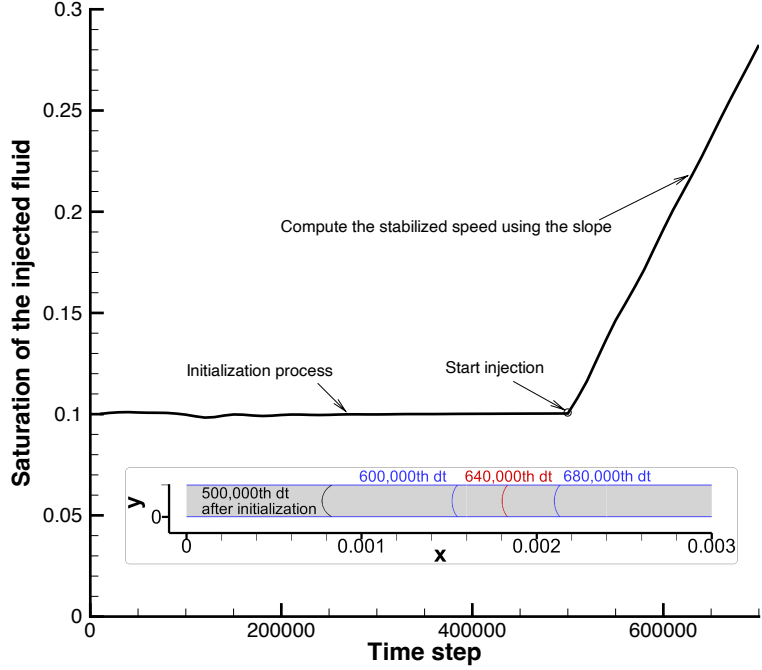

Fig. 2. A typical initialization process followed by a convergence process with a constant contact-line speed; the insert shows the corresponding initial interface as well as the deformed but stabilized interfaces.

### 4.2. Dynamic contact angle

We fist discuss the measurement of contact angle. In experiments, the dynamic contact angle can be determined via force measurement, where a solid sample is immersed in a liquid at a constant velocity and the force exerted on the solid by the liquid during immersion is recorded to determine the contact angle [15]. Additionally, the contact angle can be directly measured from the interface shape but a standard method is unavailable and, consequently, the contact angle is often obtained and reported using different techniques [3].

In our simulations, the interface shape can be clearly obtained. By assuming the interface curve as a circular arc, the curvature radius can be determined from the height $L_2$ and width $L_1 = H$ of the corresponding sector area [21], as shown in Fig. 3:

$$r = \frac{4L_2^2 + L_1^2}{8L_2}. \tag{1}$$

Then, the contact angle $\theta$ can be computed as (similar for $\theta \in (\frac{\pi}{2}, \pi]$):

$$\theta = \frac{\pi}{2} - \arctan\left(\frac{L_1/2}{r - L_2}\right), \theta \in \left[0, \frac{\pi}{2}\right]. \tag{2}$$

The above calculation is subject to noticeable uncertainty in determining $L_2$ as the exact location of contact line is hard to pinpoint and thus the error will be large as $L_2 \to 0$ (i.e., $\theta$ is close to $\pi/2$). Here, we opt to use a fitting circle that best matches the interface and determine $r$ from the fitting circle. Then, $\theta$ is computed as:

$$\theta = \arccos\left(\frac{L_1}{2r}\right), \theta \in \left[0, \frac{\pi}{2}\right] \quad \text{or} \quad \theta = \pi - \arccos\left(\frac{L_1}{2r}\right), \theta \in \left(\frac{\pi}{2}, \pi\right]. \tag{3}$$

The unit of the computed contact angle will be converted from radian to degree in the following discussion.

As an example, Fig. 3 shows the best fitting circles for the initial and stabilized interfaces of Fig. 2. The match is excellent for the initial interface while small discrepancy is visible for a circle fitting of the stabilized but deformed moving interface. This discrepancy will increase with the moving speed (or $Ca$ in general). For complicated interfaces (e.g., $Ca$=0.1206 of the group 2 in Fig. 7), the best fitting circle will be selected to match the central part of the interface as it determines the overall shape. In this case, the obtained contact angle is called the apparent contact angle in the classification of [17]. The actual interface structure is very complicated and different contact angles can be defined for various regions at different scales, as discussed in [28].

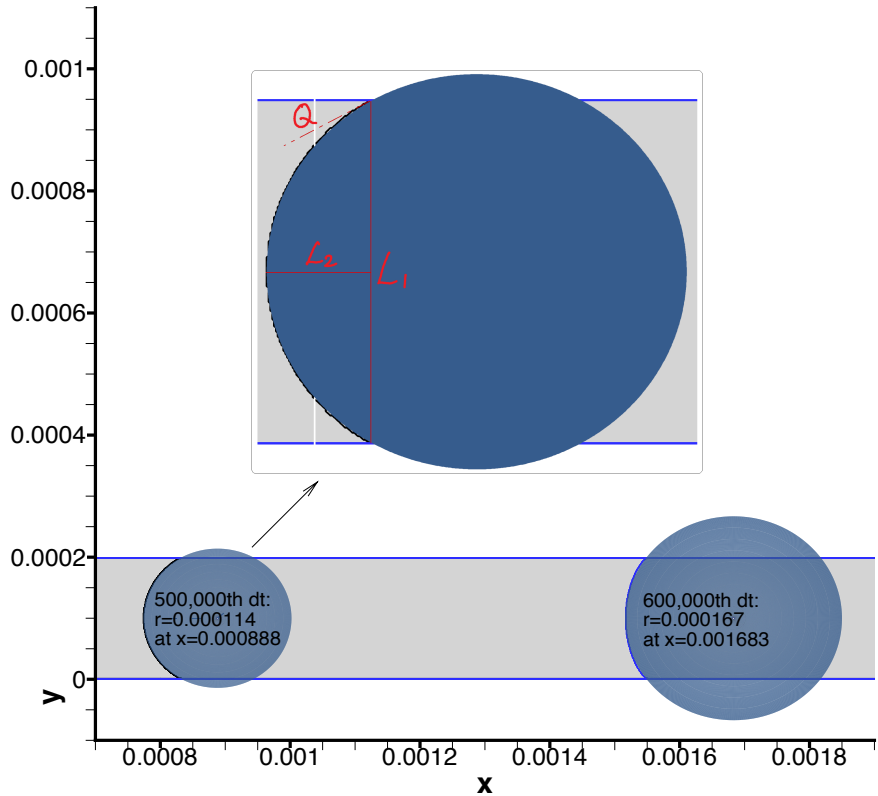

Fig. 3. Extract the curvature radii to determine the contact angles by using fitting circles for the initial static interface (the black curve fitted with a smaller circle) and a stabilized moving interface (the blue curve fitted with a larger circle) of Fig. 2.

The variation of $\theta_{dy}$ with $Ca$ is given in Fig. 4. Starting with different $\theta_{dy} = \theta_{eq}$ at $Ca = 0$, the profiles have completely different trends. The profiles with small $\theta_{eq}$ will first increase sharply and then gradually with $Ca$ while the profile with a large $\theta_{eq} \approx 152$ seems to have an ascending increase rate. As fingering occurs at very small $Ca$ for large $\theta_{eq}$, the corresponding profile is very short. For the profile with an intermediate $\theta_{eq} = 90$, the increase rate is almost constant.

Note that the cosine of the contact angle is related to the capillary force and thus the correlation between its static and dynamic values might depend on the capillary number. Many studies [5, 6, 9, 13, 14, 19] suggest that the above correlation can be presented as:

$$(\cos\theta_{\text{eq}} - \cos\theta_{\text{dy}}) = f(Ca). \tag{4}$$

In particular, the LBM simulation [5] with a viscosity ratio of 1 shows that the correlation follows a power law scaling behavior of $2\sigma(\cos\theta_{\text{eq}} - \cos\theta_{\text{dy}})/H \propto Ca^x$, where the exponent $x$ theoretically is 1 if the wall surface is ideally smooth and numerically is very close to 1 due to discretization effect. In our plotting, the coefficient $2\sigma/H$ is omitted to make the correlation dimensionless and general. Compared to the simulations of [5] with $Ca < 0.01$, our simulations cover the whole range of $Ca$ to capture the final regime with fingering at large $Ca$ and the whole range of $\theta_{\text{eq}}$. Even so, the linear scaling law is also observed in our results of Fig. 5, where all profiles starting with different $\theta_{\text{eq}}$ have almost linear fits with $x \approx 1$. This is in good agreement with the results of the macroscopic theory and the molecular dynamics simulation, as detailed in [5]. It is believed that $x$ will be smaller than 1 if the surface has roughness or chemical heterogeneity [5]. Additionally, all profiles seem to collapse onto a single line, which is also consistent with the observation of [5], where a single scaling parameter $K = 21.8$ is found valid for curves of different $\theta_{\text{eq}}$. Some deviations are visible in Fig. 5 at large $Ca$, which is because the interface is highly deformed and the error from determining $\theta_{\text{dy}}$ becomes significant.

Qualitatively, we have excellent agreement with the observations of [5], namely both show the linear scaling law that is universal for different $\theta_{\text{eq}}$. We stopped short of directly comparing the specific data because the contact angle of [5] is measured using $L_2$ that is subject to large uncertainty particularly when $\theta_{\text{dy}}$ is close to $\pi/2$ (i.e., $L_2 \to 0$) while the contact angle in our study is measured using a fitting circle to avoid $L_2$. There is no standard method for this measurement [3] but our option is necessary in the current study due to large interface deformation at large capillary numbers. We expect difference in the measured dynamic contact angle even if the interface shapes obtained by both simulations are in good match. This can be clearly seen from the highly deformed interfaces of high $Ca$ in the following figures, where the fitting circle intercepts with the walls at a location far away from the actual contact point, which indicates a completely different dynamic contact angle if the actual contact point is used for the measurement. Therefore, quantitative study of the contact angle curves becomes a matter of choice, but the above comparison shows that the obtained qualitative features are independent of the contact angle measurement and consistent between the two studies.

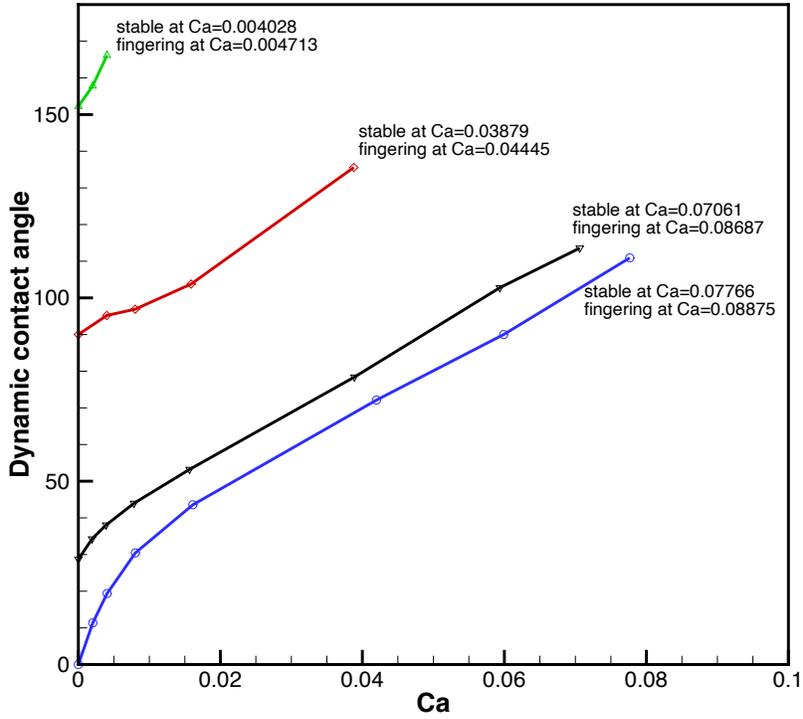

Fig. 4. Dynamic contact angle at various $Ca$ starting with different equilibrium contact angles at $Ca = 0$.

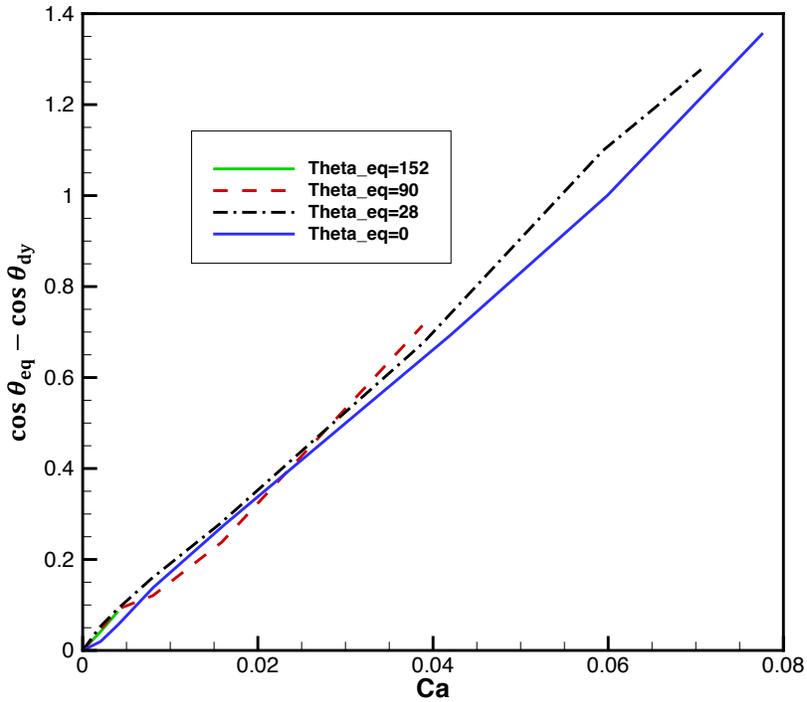

Fig. 5. Linear scaling law for the cosine of contact angle.

As mentioned before, it is hard to pinpoint the critical capillary number $Ca^*$ and the correspondingly $\theta_{dy}^*$. However, our maximum $Ca^{max}$ (and the corresponding stable $\theta_{dy}^{max}$) should be very close to the theoretical $Ca^*$ (and $\theta_{\text{dy}}^*$) as indicated by the upper bound at which fingering has been confirmed. Additionally, the results show that $Ca^{\max}$ decreases with increasing $\theta_{\text{eq}}$ but $\theta_{\text{dy}}^{\max}$ increases with $\theta_{\text{eq}}$.

For different $\theta_{\text{eq}}$, the fingering patterns are completely different. Figure 6 shows the two scenarios with $\theta_{\text{eq}} \approx 28$ and $\theta_{\text{eq}} \approx 180 - 28$, respectively. For the first scenario, the injected fluid is the wetting phase and the occurrence of fingering needs to twist the interface for a large angle by using a large $Ca$, leaving a thick film of the displaced fluid behind the interface. The resulting finger has a neck much smaller than the head. In contrast, when the injected fluid is the non-wetting phase, the static interface is already convex to the driving direction, which facilitates the fingering. In this case, the interface deformation due to the contact-line movement is very small, except leaving a very thin film of the displaced fluid behind the interface, and thus the required $Ca$ for fingering is small. The finger neck becomes thicker at larger $\theta_{\text{eq}}$. We note that the fingering pattern is similar to those previously observed in the LBM simulations [29].

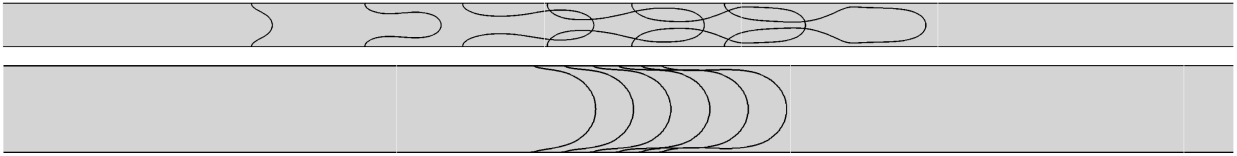

Fig. 6. Finger growing with time for the simulations with $\theta_{\text{eq}} \approx 28, Ca = 0.08687$ (top) and $\theta_{\text{eq}} \approx 180 - 28, Ca = 0.004713$ (bottom).

### 4.3. Hysteresis effect of strong complete wetting

Complete wetting (spreading) with $\theta_{\text{eq}} = 0$ can be achieved by using different settings in the LBM simulations. As an example, we adopted $\rho_w^1 = 1, \rho_w^2 = 350$ for the group 1 and $\rho_w^1 = 1, \rho_w^2 = 800$ for the group 2. The curvature radius of the static interface could be smaller than a half of the channel height due to the presence of a thin film spreading from the wetting fluid 2. As shown in the insert of Fig. 7, the fitting circle of group 1 at the initial static state of $Ca = 0$ is very close to the wall surface although a very thin film of the wetting fluid is visible. As the wettability is further enhanced in the group 2, the film thickness is increased and the fitting circle at $Ca = 0$ becomes noticeably away from the wall surface. This leads to hysteresis in the variation of $\theta_{\text{dy}}$ with $Ca$ for the group 2, where a substantial $Ca$ is required to initiate the deviation of $\theta_{\text{dy}}$ from $\theta_{\text{eq}} = 0$. Note that the surface in our study has no roughness or heterogeneity that is usually needed for the conventional hysteresis between the *static* advancing and receding contact angles at $Ca = 0$. However, the conventional hysteresis could also occur in the presence of free liquid films [4], which is consistent with our observation of a new type of hysteresis but for the *dynamic* contact angle. At a more fundamental level, the presence of fluid film in both cases is associated with the action of surface forces and thus the corresponding hystereses for both static and dynamic contact angles have the same origin.

Since the observed hysteresis might be subject to the grid resolution that should be fine enough to model the fluid film effect, we conducted the grid convergence study for the group 2 of Fig. 7. The original grid number of $8000 \times 200$ is very large making a further refinement very costly (*note*: the current $600,000 \Delta t$ in total should also be increased proportionally), we therefore coarsened the grid number to $4000 \times 100$. The comparison of stabilized interface shapes is given in Fig. 8 for two capillary numbers around the ending point of hysteresis at $Ca \approx 4.053 \times 10^{-3}$ for the group 2. As expected, the interface shape changes slightly with the grid resolution. However, the central part of interface that defines the dynamic contact angle is in good agreement. This confirms that the observed hysteresis is independent of the grid resolution. The fluid-fluid interface should be smooth in physics but is staggering in LBM simulations, which becomes worse when using a lower grid resolution. The mismatch between results of different grid resolutions mainly occurs along the staggering parts. On the other hand, we can see that the staggering effect in the current high-resolution results is already small and thus expect the change to be small should the grid number be increased.

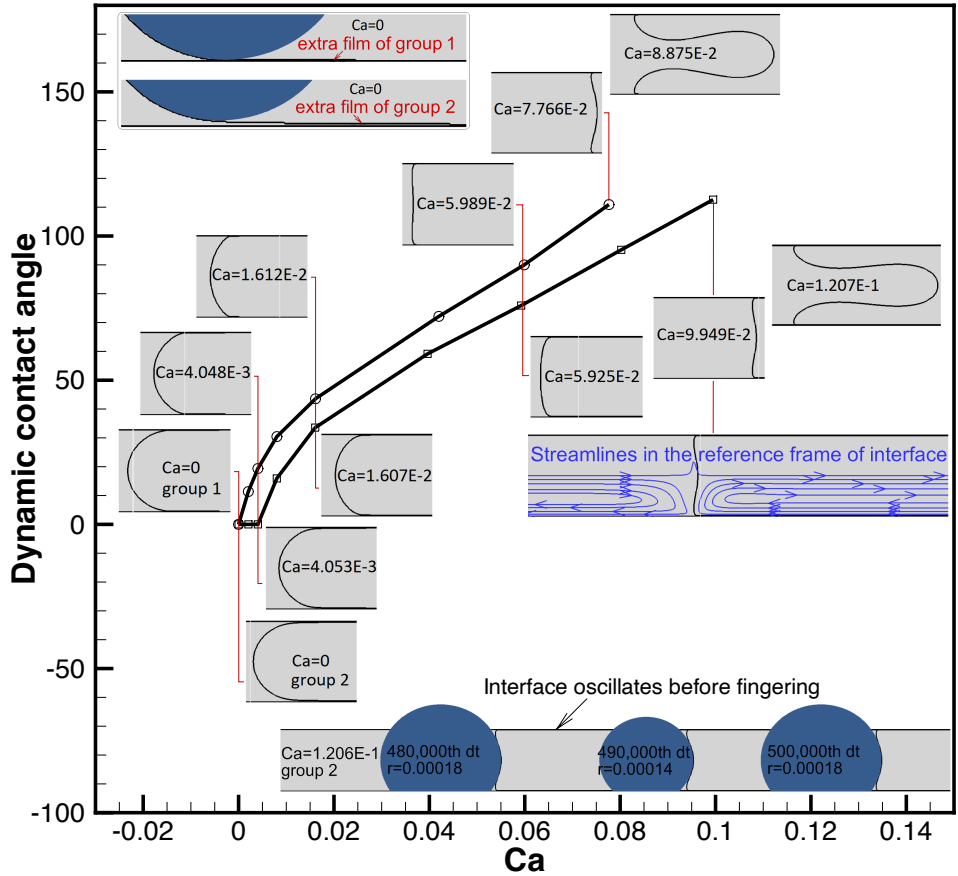

Fig. 7. Hysteresis in the evolution of dynamic contact angle between two complete wetting cases of $\theta_{eq} = 0$.

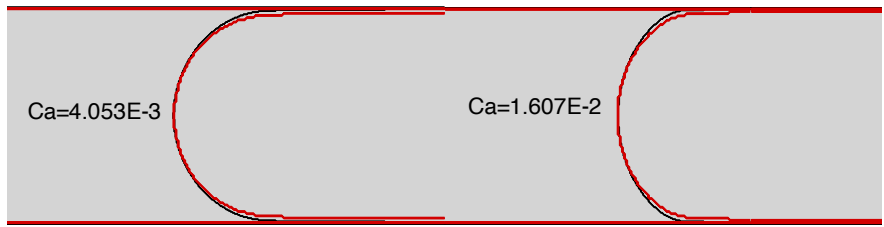

Fig. 8. Study on the grid resolution effect to the hysteresis of group 2 in Fig. 7 by using $8000 \times 200$ (black) and $4000 \times 100$ (red) grids, respectively.

Figure 7 also shows the evolution of the stabilized interface shape with an increasing $Ca$. The interface that is originally concave to the driving direction first changes to an almost flat interface with $\theta_{\text{dy}} \approx 90$ and then to an interface convex to the driving direction. With a further increase of $Ca$, the interface starts oscillating with the curvature radius going up and down, as shown in the insert for the group 2 at $Ca = 0.1206$. Similar interface oscillation phenomenon has been reported in the simulation of droplet movement at high speed [11]. We further increased the injecting speed by a factor of 7/6 and then the interface starts fingering. However, the increase of contact-line speed is negligible as the temporal variation of contact-line location indicates $Ca = 0.1207$. The evolution processes are very similar for both groups, except the difference in the required $Ca$ for each resulting interface shape due to the hysteresis effect discussed above.

The streamline pattern is also shown in the insert of Fig. 7. In the frame of reference moving together with the interface, the advancing fluid rolls over the solid boundary while the receding fluid unrolls from the solid boundary, which explains the coexistence of the slip velocity around the contact line and the traditional non-slip boundary condition for fluid away from the contact line, as usually observed in the global static frame of reference. Our streamline pattern is similar to that computed using a mean-field free-energy LBM in [6] but different from the theoretical illustrations of [16, 17]. Inside both fluids, we didn't observe the 'jet-like' structure having the bulk velocity directed from (or towards) the contact line. This 'jet-like' structure can help the theoretical analyses satisfy the velocity continuity across the fluid-fluid interface. Alternatively, the velocity continuity is satisfied in our results since the relative velocities of both fluids to the interface approach zero from different sides (e.g., $u_y \to 0_-$ and $u_y \to 0_+$ from the left and right sides of the interface, respectively). More investigations are needed to verify the structure of streamline pattern around the contact line.

It is noteworthy that as the interface has a non-zero thickness, an unbiased rule to determine the interface location is to first compare the local densities of the two components/fluids at the same spatial grid and then designate grids with $\rho_2 > \rho_1$ as being occupied by the fluid 2, and vice versa. Then, the visualization can automatically find the location of the fluid-fluid interface (i.e., a line that separates the two fluids), part of which shows the location of the film.

At the end, we would like to note that the application of the dynamic contact angle is not always straightforward. For instance, the computed capillary force using $\sigma \cos\theta_{\text{dy}}$ will be counterproductive (a dragging force) due to $\theta_{\text{dy}} > 90$ at large $Ca$ even though $\theta_{\text{eq}} < 90$ that indicates an inherent driving force. From the energy conversion point of view, the system has prestored a surface energy at the initial static state that will be released (not accumulated) as long as the wetting phase is displacing the non-wetting one, which requires the capillary force to do positive work. Interpreting the capillary force as a dragging force will lead to a negative work that is unphysical. For such a complicated interface with $\theta_{\text{eq}} < 90$ but $\theta_{\text{dy}} > 90$, the capillary force needs to be determined from the contact angle defined at a much smaller scale around the contact line, that is different from the current overall apparent contact angle, as differentiated in Section 4.2. Besides changing $\theta_{\text{eq}}$ to an appropriate $\theta_{\text{dy}}$, $\sigma$ also needs to change in computing the capillary force such that the released amount of the prestored surface energy is independent of the actual flow/releasing speed, which can be evaluated by the capillary force $\sigma \cos\theta_{\text{eq}}$ at the equilibrium state and the channel sizes as a special case of an infinitely slow releasing process. Changing both properties is also consistent with the theory that according to the force balance of the Young equation the dynamic contact angle deviating from $\theta_{\text{eq}}$ implies the change of $\sigma$ in the vicinity of contact line, which is explained using the flow circulation around the interface [17]. In a recent application [30], the calculation of capillary driving force considered only the dynamic contact angle's deviation from the static value, namely changed the released amount of surface energy, but still obtained good prediction of experimental data. This is probably because they applied the classical Poiseuille flow solution, which is valid only in the domain far away from the fluid-fluid interface, inside the whole domain and thus neglected the influence on the viscous resisting force by the flow circulation around the fluid-fluid interface (see the insert of Fig. 7). From the energy conversion point of view, their model changed the released amount of surface energy but also neglected the energy dissipation due to the circulation effect.

## 5. Conclusions

The correlation of the dynamic contact angle with the capillary number is simulated by the LBM and the profile is found to change noticeably with the equilibrium contact angle. However, the cosine of the contact angle follows a linear scaling law in the whole range of capillary numbers and the scaling lines of different equilibrium contact angles collapse onto the same one. Our observations are qualitatively consistent with the macroscopic theory and the previous simulation studies using the MD and the LBM. The upper bound of the capillary number that leads to fingering decreases significantly with an increasing equilibrium contact angle. As the interface instability also depends on the viscosity contrast, the profiles are expected to change with the viscosity contrast [16, 17]. Substantial modifications to the current LBM algorithm are needed to handle large viscosity contrasts, which will be investigated in our future work. The current study can also be extended to the flows around curved surface, which can investigate the potential dependence of the correlation on the geometry characters, as discussed in [19, 20].

For the special case of complete wetting with a zero equilibrium contact angle, hysteresis effect is observed in the dynamic contact angle variation with the capillary number when the wettability is very strong, making the wetting fluid spread over the wall surface and form an extra film. In this case, attention is needed if we want to make comparisons with the experimental study and the initial curvature radius of experiments should be provided as another property for the simulation setting of the static interface shape.

Although the contact angle can be directly measured from the interface shape, a standard method is unavailable and, consequently, the contact angle is often obtained and reported using different techniques [3]. Due to using a different measurement of the dynamic contact angle, which is necessary in the current study for highly deformed interfaces at large capillary numbers, we stopped short of making quantitative comparison with previous simulation studies. Nevertheless, the main observations of the linear scaling law even at large capillary numbers and the new hysteresis effect due to strong wettability could be interesting topics for future study.

**Data availability**

The data that support the findings of this study are available from the corresponding author upon reasonable request.

**Appendix: the LBM algorithm based on the Shan-Chen model**

The D2Q9 model [31] is used in our 2D simulations to determine the lattice velocity $\vec{e}_\alpha$ and the weight factor $\omega_\alpha$ for each $\alpha$ direction. In the Shan-Chen model [25, 26], the number of molecules of the $\sigma^{\text{th}}$ component/fluid having the lattice velocity $\vec{e}_\alpha$ at the position $\vec{x}$ and time $t$ is denoted by $f_\alpha^\sigma(\vec{x}, t)$, where $\sigma = 1, 2$ for the two immiscible fluids. At the initial state, we prescribe $f_\alpha^\sigma(\vec{x}, 0) =$ $f_\alpha^{\sigma(\text{eq})}(\rho^\sigma, \vec{u}^{\sigma(\text{eq})})$ using the initial distributions of density $\rho^\sigma(\vec{x}, 0)$ and velocity $\vec{u}^{\sigma(\text{eq})}(\vec{x}, 0) = 0$. The explicit updating algorithm of the only unknown $f_\alpha^\sigma(\vec{x}, t)$ is:

$$f_\alpha^\sigma(\vec{x} + \Delta t \vec{e}_\alpha, t + \Delta t) = f_\alpha^\sigma(\vec{x}, t) + \frac{f_\alpha^{\sigma(\text{eq})}(\vec{x}, t) - f_\alpha^\sigma(\vec{x}, t)}{\tau^\sigma}, \tag{A1}$$

where the normalized relaxation time $\tau^\sigma$ is determined according to the kinematic viscosity of the $\sigma^{\text{th}}$ fluid and the equilibrium distribution function is defined as:

$$f_\alpha^{\sigma(\text{eq})} = \rho^\sigma \omega_\alpha \left[1 + \frac{3}{c^2} \vec{e}_\alpha \cdot \vec{u}^{\sigma(\text{eq})} + \frac{9}{2c^4} [\vec{e}_\alpha \cdot \vec{u}^{\sigma(\text{eq})}]^2 - \frac{3}{2c^2} \vec{u}^{\sigma(\text{eq})} \cdot \vec{u}^{\sigma(\text{eq})}\right], \tag{A2}$$

where $c = \Delta x / \Delta t$, $\rho^\sigma = \sum_\alpha f_\alpha^\sigma$. The equilibrium velocity $\vec{u}^{\sigma(\text{eq})}$ is defined as follows:

$$\vec{u}^{\sigma(\text{eq})} = \frac{\rho^\sigma \vec{u}' + \tau^\sigma \vec{F}^\sigma}{\rho^\sigma}, \tag{A3}$$

where $\vec{F}^\sigma$ is related to the total volume force, including the fluid-fluid interaction $\vec{F}^{\sigma,1}$, fluid-solid interaction $\vec{F}^{\sigma,2}$ and external force $\vec{F}^{\sigma,3} = \Delta t \rho^\sigma \vec{g}$ in general [21]. The auxiliary velocity $\vec{u}'$ is defined as follows to conserve the total momentum of the two fluids:

$$\vec{u}' = \frac{\sum_\sigma \frac{1}{\tau^\sigma} \sum_\alpha \vec{e}_\alpha f_\alpha^\sigma}{\sum_\sigma \frac{1}{\tau^\sigma} \sum_\alpha f_\alpha^\sigma}. \tag{A4}$$

The overall flow velocity $\vec{u}$ of the two fluids is equal to the mean velocity before and after implementing the force term:

$$\vec{u} = \frac{\sum_\sigma \sum_\alpha \vec{e}_\alpha f_\alpha^\sigma + 0.5 \sum_\sigma \vec{F}^\sigma}{\sum_\sigma \rho^\sigma}. \tag{A5}$$

The interaction force exerted on the fluid $\sigma$ by its surrounding fluid $\sigma' \neq \sigma$ is computed as follows:

$$\vec{F}^{\sigma,1}(\vec{x}, t) = -G \rho^\sigma(\vec{x}, t) \sum_{\alpha \in \text{fluid}} \omega_\alpha \rho^{\sigma'}(\vec{x} + \Delta t \vec{e}_\alpha, t) \vec{e}_\alpha, \tag{A6}$$

where the summation is over all $\alpha$ directions that point towards neighboring fluid grids. The fluid-solid interaction force is similarly computed:

$$\vec{F}^{\sigma,2}(\vec{x}, t) = -G \rho^\sigma(\vec{x}, t) \sum_{\alpha \in \text{wall}} \omega_\alpha \rho_w^{\sigma'} \vec{e}_\alpha, \tag{A7}$$

where $\rho_w^{\sigma'=1}$ and $\rho_w^{\sigma'=2}$ are two constant effective wall densities for the two fluids, respectively, and can be appropriately set to regulate the equilibrium contact angle [27]. Correspondingly, the pressure after considering the fluid-fluid interaction becomes:

$$p = \frac{c^2}{3}(\rho^1 + \rho^2 + G\rho^1\rho^2). \tag{A8}$$